\documentclass[12pt,tightenlines,
    onecolumn,
    preprintnumbers,
    amsmath,
    amssymb,
    prd,
    showpacs,
    showkeys
]{revtex4}
\usepackage{color}
\usepackage{graphicx}%
\usepackage{amsmath}
\usepackage[normalem,normalbf]{ulem}
\usepackage{amssymb}
\usepackage{bm}
\usepackage{enumerate}

\newcommand{\nn}{\nonumber}

\begin{document}
\title{Interacting scalar field theory in $\kappa$-Minkowski spacetime}
\author{Chaiho Rim}
\email{rim@chonbuk.ac.kr}
\affiliation{
Department of Physics and Research Institute
of Physics and Chemistry, Chonbuk National University,
Jeonju 561-756, Korea
}%
\preprint{arXiv:0802.3793[hep-th]}
\preprint{2008/Feb/26}

\bigskip
\begin{abstract}
We construct an complex scalar field theory 
in $\kappa$-Minkowksi spacetime,
which respects $\kappa$-deformed Poincar\'e symmetry.
One-loop calculation shows that the theory is finite 
and needs finite renormalization to be compatible 
with the $\kappa \to \infty$ limit.
The loop result also has an imaginary valued correction
due to the complex poles present in the propagator.

\end{abstract}
\maketitle

\section{Introduction}

Poincar\'e symmetry has been a primary geometric notion 
for the Minkowski spacetime (MST) 
and played the guiding role of 
construction of quantum field theory. 
At the short distance of Planck length scale, however, 
the spacetime itself may change its concept
due to the quantum gravity effect.
On this purpose, Poincar\'e algebra in momentum space 
is deformed~\cite{kappaP} 
and a new scale parameter $\kappa$ is 
introduced, which will be an order of Planck length.   
The $\kappa$-deformed Poincar\'e algebra (KPA)
can have many different forms. 
Based on bicrossproduct basis~\cite{majid} 
where the four momenta are commuting each other
but the boost relation is deformed,
the dual picture of the KPA is realized
in terms of non-commuting spacetime~\cite{majid,zakr}.
This non-commuting spacetime is called
$\kappa$-Minkowski spacetime (KMST),
where the rotational symmetry is preserved
but time and space coordinates
do not commute each other,
\begin{equation}
~[\hat x^0, \hat x^i] = \frac{i}{\kappa} \hat x^i\,,\qquad
 ~[\hat x^i, \hat  x^j]=0\,,\qquad i,j=1,2,3\,.
\end{equation}

The Planck scale parameter $\kappa$ 
has the role of a deformation parameter. 
When $\kappa$ approaches infinity,
the deformed Poincar\'{e} algebra in momentum space 
reduces to the ordinary Poincar\'e algebra 
and therefore, the ordinary Poincar\'{e} symmetry
is recovered in Minkowski spacetime.
The $\kappa$-deformed realization 
implies that the special relativity is deformed 
and the energy momentum relation also has a new form.
This will result in a change 
of the group velocity of photon. 
In this respect, KPA implies the doubly special relativity~\cite{doubly} 
and the deformation parameter $\kappa$
reflect the Planck scale physics.

After the appearance of the KPA, 
it is soon realized that
the differential structure of the KMST of 4 spacetime
dimension is not realized in 4 dimensional spacetime
but needs to be constructed
in 5 dimensional spacetime~\cite{sitarz,gonera}.
This reminds of the Snyder's approach 
where non-commutative coordinates are realized in 
4 dimensional De Sitter space \cite{snyder}.
The differential calculus is realized in 
exponential operator 
$\{e^{-i p \cdot \hat x}\}$ 
with an appropriate ordering of $\hat t$ and 
$\hat {\bf x}$.
\begin{equation}
d \{e^{-i p \cdot \hat x}\} \equiv 
\hat \tau^A \partial_A \{e^{-i p \cdot \hat x}\}
\end{equation} 
where $A=0,1,2,3,5$ and $\hat \tau^A$ 
is the differential element, 
$\hat \tau^\mu =d \hat x^\mu$ 
with $\mu = 0,1,2,3$ 
and $\hat \tau^5$ is 
the new differential element. 
The momentum realization of the derivatives
is given as 
\begin{equation}
\partial_A \{e^{-i p \cdot \hat x}\} 
=\chi_A (p) \{e^{-i p \cdot \hat x}\} 
\end{equation}  
and $\chi_\mu(p)$ behaves as a 4-vector element
and $\chi_5$ as an invariant in KPA~\cite{KRY-b}, 
\begin{align}
&\left[ N_i, \chi_j\right] 
= i \delta_{ij} \chi_0\,,\qquad
\left[ M_i, \chi_j\right] 
= i\epsilon_{ijk} \chi_k 
\nn\\
&\left[ N_i, \chi_0\right] 
= i \chi_i \,,\qquad\quad   
\left[ M_i, \chi_0\right] = 0\,
\nn\\
&\left[ N_i, \chi_5\right] 
= 0=\left[ M_i, \chi_5\right] \,.
\label{K-4vector}
\end{align}
Here $i,j=1,2,3$ and $M_i$ and $N_i$ are 
the rotation and boost generators of KPA,
respectively.
It is worth to mention 
that the corresponding derivative is realized
in the  $4-$dimensional De Sitter space:
\begin{equation}
({\cal P}_0)^2-({\cal P}_i)^2 -({\cal P}_5)^2 = -\kappa^2\,,
\end{equation}
with ${\cal P}_\mu = \chi_\mu$ 
and ${\cal P}_5 = 4\chi_5 -\kappa$.

The invariant property of $\chi_5$ 
leads one to construct 4-dimensional system 
without invoking the fifth dimensional 
tangential direction 
if one requires physical system to respect
the $\kappa$-deformed Poincar\'{e}  symmetry (KPS).
Based on this 4-dimensional differential structure,
one requires the on-shell condition to be 
\begin{equation}
\label{on-shell}
\chi_\mu \chi^\mu =m^2
\end{equation}
where $m$ is the particle mass
and constructs the scalar field theory 
with KPS~\cite{kosinski,KRY-a,KRY-b}.
Still, an interacting (field) theory 
needs more elaboration since 
it is not clear how to construct the many particle
states from the right choice of vacuum since 
the many particle states constructed so far 
show the non-local nature. 
(See for example \cite{kosinski,DLW} and references there in).

To understand the physical effects of the 
$\kappa$-deformation, one may study 
black-body
radiation~\cite{KRY-b} and Casimir energy~\cite{KRY-c}
using the mass-shell condition only, 
which uses essentially the free field theory only. 
It turns out that 
the thermal energy of the blackbody radiation 
due to the massless mode of the KMST 
($m^2=0$ in (\ref{on-shell}))
reduces to the Stephan-Boltzmann law 
(proportional to $T^4$) 
when $\kappa \to \infty$ limit is taken 
if one takes care of ordinary modes (OM)
from the mass-shell condition (\ref{on-shell}) 
which reduces to the one from Einstein's special relativity.

In the asymmetric ordering, 
the ordinary massless mode
is explicitly given as
\begin{equation}
\label{asym-OM} 
\Omega_{\bf p}^+
	= - \kappa \log(1-\frac{|{\bf p}|}{\kappa})\,,\qquad 
\Omega_{\bf p}^-
=  \kappa\log(1+ \frac{|{\bf p}|}{\kappa}) \,.
\end{equation}
In fact, the mass-shell condition (\ref{on-shell}) 
also allows high momentum mode (HM) 
which exist only when its momentum is greater 
than $\kappa$.
\begin{equation}
\label{asym-HM} 
\Omega_{\bf p}^H
	= - \kappa \log(\frac{|{\bf p}|}{\kappa} -1)\,.
\end{equation}
It is shown in  \cite{KRY-b} that
the Stephan-Boltzmann law would be spoiled 
if the HM were to be included 
in the thermal distribution, 
whose contribution 
turns out to be proportional to $T^2$ or $T^3$ 
depending on how one treats 
the negative energy part of HM. 
Thus, one needs to eliminate the HM from the 
on-shell. The same thing applies to the symmetric 
ordering case. 

On the other hand, the study of Casimir energy 
on a spherical shell shows that 
in the asymmetric ordering case the vacuum 
can break particle and anti-particle symmetry 
at Planck scale:
The Casimir energy of the negative mode 
(anti-particle) in (\ref{asym-OM}) 
is not the same as the one due to the positive mode 
(particle) if the HM is not included.
Thus, if one requires the vacuum 
to respect the particle and antiparticle symmetry 
at Planck scale, one cannot adopt the asymmetric 
ordering dispersion relation. 
This reasoning forces us to adopt 
the symmetric ordering only to have 
the particle and anti-particle symmetry
at the Planck scale. 

In this paper we are going to construct 
an interacting complex scalar field theory 
imposing the KPS. 
We present the essential element for 
the free field theory 
in Sec.~\ref{sec:2}, 
and construct its interacting scalar field theory 
in Sec.~\ref{sec:3}. We evaluate 
the one-loop correction of propagator
in Sec.~\ref{sec:4} and one-loop correction 
of vertex in Sec.~\ref{sec:5}. 
Sec.~\ref{sec:6} is the summary and discussion.

\section{Free scalar field theory} 
\label{sec:2}

To construct the free field theory with KSP
one defines a field variable in momentum space,
\begin{equation}
\phi(x) \equiv
\int \frac{d^4 p}{(2\pi)^4}  \,
e^{-i p \cdot  x}\,\varphi (p)\,.
\end{equation}
Here both  the coordinate variable $x$  
and momenta $p$ are treated as commuting variables.
The non-commuting nature of KMST is encoded
in $\kappa$-$\ast$ product between field variables:
The product of exponential element is required to
satisfy the composition rule \cite{FGN}
\begin{equation}
 e^{-i p \cdot  x}\ast e^{-i q
\cdot x} = e^{-i v (p, q) \cdot x }\,.
\end{equation}
In this paper, we will adopt 
the composition law 
corresponding to the symmetric ordering;
\begin{equation}
v (p, q)
= ( p^0+q^0, {\bf p} e^{-\frac{q^0}{2\kappa}}+ 
{\bf q} e^{\frac{p^0}{2\kappa}}) \,.
\end{equation}
The homomorphism of the product of field variables 
reproduces the KMST effect 
and this way, one can avoid 
various conceptual difficulties of non-commuting spacetime geometry.

The KPS is the guiding principle to construct the field theory and is applied to the free scalar action explicitly
in~\cite{KRY-b}.
The free analogue of massive complex scalar theory is
given as
\begin{equation}
\label{action-coor}
S_2= \int d^4 x \, \phi^c(x)\ast
 \left[-\partial_\mu \ast   \, \partial^\mu \ast
  - m^2 \right] \, \phi(x)\,.
\end{equation}
$\phi^{c}(x)$ is the conjugate of the scalar field
and is expressed just as the complex conjugate of 
the field in this symmetric ordering case:
\begin{equation}
\phi^{c}(x) =  \int_{ p} e^{i p\cdot x }\, \varphi^\dagger (p)\,,
\end{equation}

In momentum space, 
the action in (\ref {action-coor})
is given as
\begin{equation}
\label{action-momentum}
S_2= \int \frac{ d^4 p}{(2\pi)^4} \, 
e^{\alpha  p^0}
\varphi^\dagger(p) 
\Big( \chi^\mu \chi_\mu (p) -m^2 \Big)
 \, \varphi(p) 
\end{equation}
where $\kappa$-Poincar\'e invariance sets 
$\alpha=3/(2\kappa)$ 
(see below (\ref{K-measure})).
Explicit form of $\chi_\mu$ is given as 
\begin{align}\label{chi}
& \chi_0(p) 
	= \kappa\left[\sinh \frac{p_0}{\kappa} 
		+\frac{{\bf p}^2}{2\kappa^2}\right]\,,\qquad 
		\chi_i(p)= p_i e^{\frac{p_0}{2\kappa}}
\,. \nn
\end{align}
$\chi_\mu$ is the 4-vector (\ref{K-4vector})
and  $\chi^\mu \chi_\mu (p)$ and $\chi_5$ are invariants in KPA 
\begin{align}
\chi^\mu \chi_\mu (p) & 
= \chi^\mu\,\chi_\mu (-p)
= M^2_s(p)\left(1+\frac{M^2_s (p)}
{4\kappa^2}\right)
\\
\chi_5 &= - \frac{M_s^2(p)}{8\kappa} 
\end{align}
where  $M_s^2(p)$ is the Casimir invariant  
\begin{equation}
M_\kappa^2(p)= M_\kappa ^2(-p)
=\left(2 \kappa \sinh \frac{p_0}{2 \kappa }\right)^2- {\bf p}^2\,.
\end{equation} 
One notes that the the integration measure 
given in (\ref{action-momentum}) 
is invariant under the KPS:
\begin{equation}
\label{K-measure}
d^4 p\, e^{\alpha p^0 }  = \frac{\kappa}4 \,
\frac{ d\chi_0 d\chi_1 d\chi_2 d\chi_3} {\chi_5}\,.
 \end{equation} 
 
Let's introduce a notation 
for the propagator function
$\Delta^{-1}(p) = { \chi_\mu \chi^\mu - m^2} $
which is explicitly written as
\begin{align}
\label{feynman}
\Delta (p) &= 
\frac{4\kappa^2}
{\Big(2 \kappa^2 \cosh(p^0/\kappa) 
	- {\bf p}^2  - m_\kappa^2  \Big)
\Big( 2 \kappa^2 \cosh(p^0/\kappa) 
	- {\bf p}^2+  m_\kappa^2  \Big)}
\\ \nn
m_\kappa^2  &= 2 \kappa^2 \sqrt{1 + m^2/\kappa^2}\,.
\end{align} 
The on-shell dispersion relation is given as 
$\Delta^{-1} (p) =0$;  
\begin{align}
\label{OMdispersion}
2 \kappa^2 \cosh(p^0/\kappa)  &= {\bf p}^2 +  m_\kappa^2 
\\
\label{TMdispersion}
2 \kappa^2 \cosh(p^0/\kappa)  &= {\bf p}^2  - m_\kappa^2\,. 
\end{align}
The dispersion relation in~(\ref{OMdispersion}) 
corresponds to the ordinary mode (OM) 
since this reduces to 
the ordinary particle and antiparticle  
dispersion relation as $\kappa \to \infty$.
The second one~(\ref{TMdispersion})
corresponds to the tachyon mode 
since the mode is obtained 
by putting $m_\kappa \to i m_\kappa$ in~(\ref{OMdispersion}).
The tachyon mode,
when its momentum is sufficiently 
large ${\bf p}^2  - m_\kappa^2>0$,
becomes a real mode corresponding to the 
high momentum mode (HM).
This HM should not be included in on-shell mode
since HM will spoil the blackbody radiation law 
at $\kappa \to \infty$ limit~\cite{KRY-b}. 

The propagator function 
$\Delta(p)$ has the periodic property
\begin{equation}
\label{periodicity}
 \Delta(p_0+ i 2\kappa \pi, {\bf p} )
=\Delta(p_0, {\bf p})
\end{equation}
and thus possesses an infinite number of
poles on the complex plane of $p^0$.
It is convenient for later use 
to separate the OM and TM contribution, 
each satisfying the periodicity relation (\ref{periodicity});
\begin{align}
\label{OT}
\Delta (p) 
&= \Delta_{\rm P} (p) - \Delta_{\rm T} (p)
\nn\\
\Delta_{\rm P} (p) 
&= \frac{2\kappa^2 /m_\kappa^2}
{2 \kappa^2 \cosh(p^0/\kappa) 
- {\bf p}^2  - m_\kappa^2 }	
\\
\Delta_{\rm T} (p) 
&= \frac{2\kappa^2/m_\kappa^2}
		{2 \kappa^2 \cosh(p^0/\kappa) 
		- {\bf p}^2+  m_\kappa^2 }\,.
\nn
\end{align} 

\section{Interacting Scalar field theory} 
\label{sec:3} 

We will assume there is one complex 
scalar field in this paper.
Extension to many fields 
is straight-forward. 
To find an interaction which respects KPS, 
one notices that KPS is 
preserved in the $\kappa$-$\ast$ product interaction  
\begin{equation}
\int d^4 x \phi^c_2 (x) \ast \phi_2 (x)
\end{equation}
where $\phi_2(x)$ represents composite two fields.
There are two ways to represent $\phi_2$;
\begin{align}
\phi_2^{(A)}(x) &=  \phi(x) \phi(x) 
= \int \frac{d^4 p}{(2\pi)^4} e^{-i px} 
\varphi_2^{(A)} (p)
\nn\\
\phi_2^{(B)}(x) &= \phi^c(x)  \phi(x) 
= \int \frac{d^4 p}{(2\pi)^4} e^{-i px} 
\varphi_2^{(B)} (p)
\end{align}
where 
\begin{align}
\varphi_2^{(A)} (p)
&=\int \frac{d^4 q}{(2\pi)^4} \varphi_2(p-q) \varphi_2(q)
\nn\\
\varphi_2^{(B)} (p)
&= \int \frac{d^4 q}{(2\pi)^4} 
\varphi_2^\dagger(p-q) \varphi_2(q)\,.
\end{align}
This allows two types of interactions. 
\begin{align}
S_4^{(A)} 
\label{A-type}
&= \lambda_A 
\int \frac{d^4 p } {(2\pi)^{4}}  
e^{\alpha p^0}  
\varphi_2^{(A)\dagger} (p) \varphi_2^{(A)} (p) 
\\ 
&= \lambda_A  \int \frac{d^4 p_1 d^4 p_2 d^4 p_3 d^4 p_4 } {(2\pi)^{12}}  
e^{\alpha (p_1^0 + p_2^0)} 
\varphi^\dagger (p_1) \varphi^\dagger (p_2)  
\varphi (p_3) \varphi (p_4)   
\delta^{(4)}(p_1+p_2 -p_3 -p_4 )  
\nn\\
\label{B-type}
S_4^{(B)} 
&= \lambda_A 
\int \frac{d^4 p } {(2\pi)^{4}}  
e^{\alpha p^0}  
\varphi_2^{(A)\dagger} (p) \varphi_2^{(A)} (p) 
\\ 
&=  
\int \frac {d^4 p_1 d^4 p_2 d^4 p_3 d^4 p_4 } {(2\pi)^{12}}  
\cosh \left( \frac{ \alpha(p_1^0 - p_2^0)} {2} \right) 
\cosh \left( \frac{\alpha (p_3^0 - p_4^0)} {2} \right) 
\nn\\
& ~~~~~~~~~~~~~ 
\varphi^\dagger (p_1)  
\varphi^\dagger (p_2)  
\varphi (p_3) \varphi (p_4)  
\delta^{(4)}(p_1+p_2 -p_3 -p_4 )  
\nn
\end{align}
where the bosonic permutation symmetry 
of the scalar field is used in the last identity.
It turns out that the B-type interaction, 
however, spoils the KPS 
after loop correction. 
Thus we will consider A-type interaction only 
(\ref{A-type}). 

\begin{figure}[hp]
\begin{center}
\includegraphics[width=.45\linewidth,origin=tl]{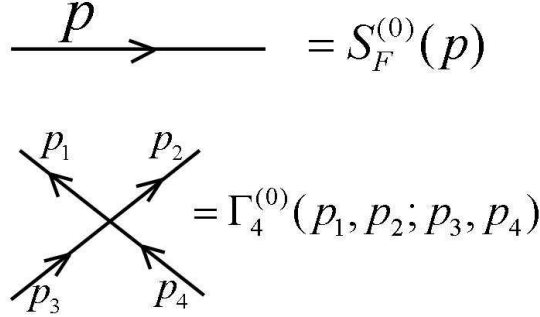}
\end{center}
\caption{Feynman rule for two and four point function}
\label{fig:feynman}
\end{figure}

Our action is written as 
\begin{align}
S&= \int d^4 x \, \left( \phi^c(x)\ast
 \left[-\partial_\mu \ast   \, \partial^\mu \ast
  - m^2 \right] \, \phi(x)
  - \frac{\lambda}4 \,
  \phi_2^c(x)\ast \phi_2(x)\right)
\nn\\
&= \int \frac{d^4 p } {(2\pi)^{4}}  
e^{\alpha p^0} \varphi^\dagger(p) \varphi(p) 
\Delta^{-1}(p) 
\nn\\
&~~~~~~~
- \frac{\lambda}4 
\int \frac{d^4 p_1 d^4 p_2 d^4 p_3 d^4 p_4 } {(2\pi)^{12}}  
e^{\alpha(p_1^0 + p_2^0) } 
\varphi^\dagger (p_1) \varphi^\dagger (p_2)  
\varphi (p_3) \varphi (p_4)   
\delta^{(4)}(p_1+p_2 -p_3 -p_4 )  
\end{align}
From this action the Feynman rule follows.
The propagator is given as
(see Fig.~\ref{fig:feynman} for notation)
\begin{equation}
S_F^{(0)}(p_1,p_2) 
= S_F(p_1) \,\, \delta^{(4)} (p_1-p_2) 
\end{equation}
where $S_F(p) ={ e^{-\alpha p^0}} {\Delta(p)}$.
Four point vertex is given as 
$\Gamma_4^{(0)} (p_1, p_2; p_3, p_4)$  where 
\begin{equation}
\Gamma_4^{(0)}(p_1, p_2; p_3, p_4)
= \lambda \,\,e^{\alpha(p^0_1 + p_0^2)}\,\,
(2\pi)^4 \delta^{(4)}(p_1+p_2 -p_3 -p_4 ) \,.
\end{equation}

\section{One loop correction of the propagator} 
\label{sec:4} 

One loop correction of the propagator is given as 
\begin{align}
S_F^{(1)}(p) &= \lambda \int 
 \frac{d^4 q } {(2\pi)^{4}}  
 \,\,e^{\alpha(p^0 + q_0)}\,\,
{ e^{-\alpha q^0 }}{\Delta(q)}
 \equiv 
 e^{\alpha p^0} \Delta \Gamma_2^{(1)}(p)\,
 \end{align}
where $\Delta \Gamma_2^{(1)}(p) $
is independent of the external momentum $p$
\begin{equation}
\Delta \Gamma_2^{(1)}(p) 
= \lambda \int 
 \frac{d^4 q } {(2\pi)^{4}}  \,\,{\Delta(q)}\,.
\end{equation}
 Using the explicit form of $\Delta(p) $ in (\ref{feynman})
one may put this two point function correction as 
\begin{align}
\Delta \Gamma_2^{(1)}(p) 
&= \lambda \int \frac{d^4 q}  {(2\pi)^{4}} 
\left\{ \Delta_{\rm P}(q) - \Delta_{\rm T}(q) 
\right\}
\nn\\
&= \lambda\frac{\kappa}{m_\kappa^2} 
\int 
 \frac{d^3 {\bf q} }  {(2\pi)^{3}} 
\int_0^\infty  \frac{dt}{2\pi}  
\left\{  \frac1{(t -\beta)(t-1/\beta) }
- \frac1{(t -\gamma)(t-1/\gamma)}\right\}\,
\end{align}
where $t = e^{q^0/\kappa}$ is used in the integration and
\begin{equation}
\beta + \frac1\beta 
= 2 a    \,,\quad
a= \frac{ {\bf q}^2 + m_\kappa^2}{2\kappa^2} \,;\qquad
\gamma  + \frac1\gamma = 2 b  \,,\quad
b= \frac{{\bf q}^2 - m_\kappa^2 }{2\kappa^2} \,.
\end{equation} 
Here $a >1$ and  $\beta $ and $1/\beta$  are 
positive real for the whole range 
of the 3-momentum integration.
Therefore, the integrand has 
two simple poles at $t=\beta$ and $1/\beta$
where we set 
\begin{align}
\beta &= a + \sqrt{a^2-1}  \,,\quad
1/\beta = a - \sqrt{a^2-1}\,.
\end{align} 
To avoid this singularity, 
one employs 
the small $\epsilon$ prescription 
so that the pole lies at 
$t= \beta - i \epsilon, 1/\beta + i \epsilon$. 
The $\epsilon$ prescription  
($t= \beta- i \epsilon$, 
$t= 1/\beta + i \epsilon$) 
amounts to put 
($p^0 = \sqrt{m^2 + {\bf p}^2} -i \epsilon$, 
$p^0 = -\sqrt{m^2 + {\bf p}^2} +i \epsilon$)
in the ordinary field theory 
when $\kappa \to \infty$.

\begin{figure}[hp]
\begin{center} 
\includegraphics[width=90mm]{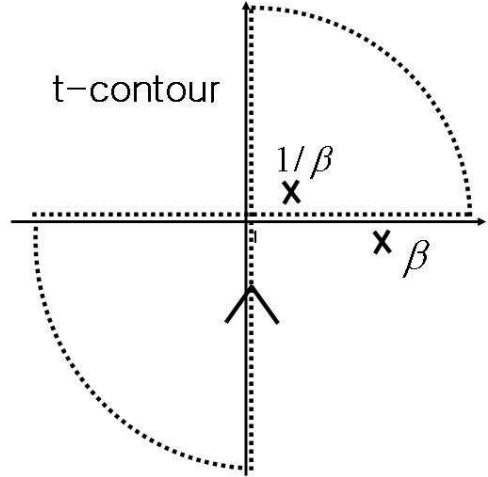}
\end{center} 
\caption{contour integration of $t$.}
\label{fig:contour1}
\end{figure}

After this prescription, 
one can evaluate the integration 
using the contour given in 
Fig.~(\ref{fig:contour1}).
The contribution over the quarter circles 
at infinity is neglected since it is canceled 
by the tachyon contribution.
This prescription results in the integration
\begin{align}
\label{IP}
{\cal I}_P (\beta)
& =
\int_0^\infty  \frac{dt}{2\pi}  
\frac1{(t -\beta +i \epsilon)(t-1/\beta -i\epsilon)} 
={\cal I}_P^{(W)} (\beta)
+{\cal I}_P^{(C)} (\beta)
\nn\\
{\cal I}_P^{(W)} (\beta)
&=\frac{-i}{\beta -1/\beta } 
\nn\\
{\cal I}_P^{(C)} (\beta)
&= i
\int_0^\infty  \frac{d\tau}{2\pi}  
\frac1{(i\tau -\beta )
(i\tau-1/\beta)} 
=\frac{-1}{\beta -1/\beta } 
\left( \frac{\ln(\beta^2)}{2\pi}\right)
\end{align}
where ${\cal I}_P^{(W)} (\beta)$ 
is the pole contribution and 
${\cal I}_P^{(C)} (\beta)$ is the integrated 
value along the imaginary axis.  

\begin{figure}[hptb]
\begin{center}
\includegraphics[width=90mm]{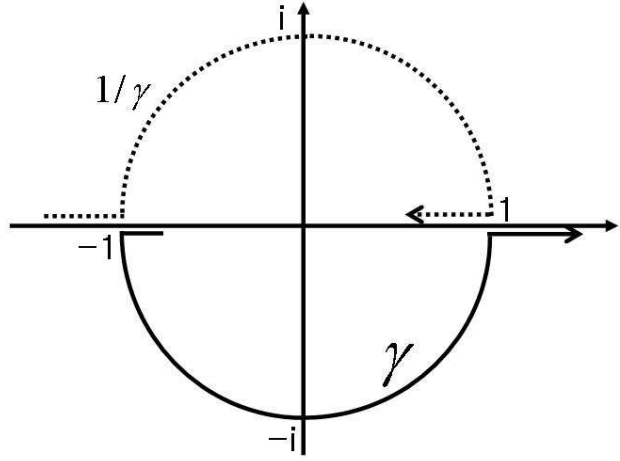}
\end{center}
\caption{Flow of $\gamma$ and $1/\gamma$ 
during the 3-momentum integration when magnitude of momentum increases.}
\label{fig:g-flow}
\end{figure}

About the tachyon contribution, 
the real part of $\gamma$ changes its sign 
depending on the 3-momentum range. 
In addition, $\gamma$ becomes complex 
when $0< b^2<1$.  
Noting that $\gamma$ and $1/\gamma$ poles 
correspond to the tachyon poles 
($m_\kappa \to i m_\kappa$ from the 
$\beta $ and $1/\beta$ pole)
one can prescribe 
$\gamma$ to lie on the lower half plane 
of the complex $t$-plane 
for the whole range of the 3-momentum
and $1/\gamma$ on the upper half plane.
\begin{eqnarray}
\label{gamma-pole}
\gamma &=& \left\{ 
\begin{array}{ll}
b + \sqrt{b^2 -1} -i\epsilon 
~~~~&{\rm when}~~b^2>1 \\
b -i \sqrt{1-b^2} 
\equiv e^{-i \Theta_\gamma} 
~~~~&{\rm when}~~-1<b<1 
\end{array} \right.
\nn\\
\frac1\gamma &=&
\left\{ 
\begin{array}{ll}
b - \sqrt{b^2 -1} + i\epsilon 
~~~~~~&{\rm when}~~b^2>1 \\
b +i  \sqrt{1-b^2 } 
\equiv e^{i \Theta_\gamma} 
~~~~~~&{\rm when}~~-1<b<1 
\end{array}
\right. 
\end{eqnarray}
where $\cos \Theta_\gamma = b~$ and 
$~0 < \Theta_\gamma <\pi$. 
The schematic flow is 
seen in the Fig.~\ref{fig:g-flow}.

This prescription results in the 
$t$-integration  
\begin{align}
\label{IT}
{\cal I}_T (\gamma) &=
\int_0^\infty  \frac{dt}{2\pi}  
\frac1{(t -\gamma +i\epsilon)(t-1/\gamma-i\epsilon)} 
={\cal I}_T ^{(W)}  (\gamma) 
+ 
{\cal I}_T ^{(C)} (\gamma)  
\nn\\
{\cal I}_T^{(W)}  (\gamma)  
&=\frac{-i}{\gamma - 1/\gamma} \,\, \theta(b)
\nn\\
{\cal I}_T^{(C)} (\gamma) 
&= i
\int_0^\infty  \frac{d \tau}{2\pi}  
\frac1{(i\tau -\gamma)
(i\tau-1/\gamma)} 
=\frac{-1}{\gamma - 1/\gamma} 
\left( \frac{\ln\gamma^2}{2\pi}\right) 
\end{align}
where
$\theta(b) $ is the Heavyside step function
so that the pole contribution is not vanishing
only when $b>1$. 

After the $t$-integration, 
one is left with the form,
\begin{equation}
\Delta \Gamma_2^{(1)}(p) 
=\lambda \kappa^2  \,  \frac{ F(x)}x\,.
\end{equation} 
Here, $F(x)$ is 
a function of $x\equiv {m_\kappa^2}/{(2\kappa^2)} 
= \sqrt{1 +m^2/\kappa^2}$
with $x \ge 1$
and is given 
in terms of  the 3-momentum integration:
\begin{align}
F(x)&= -\frac{\sqrt{2}}{\pi^2} 
\int_0^\infty r^2 dr 
\left(
\frac1{B-1/B } 
\Big(i + \frac{\ln(B^2)}{2\pi} \Big) 
- \frac1{C-1/C} 
\Big(i \theta( r -r_2) 
+ \frac{\ln(C^2)}{2\pi} \Big)
\right)
\\
B&= r^2 +x +\sqrt{(r^2+x)^2-1} 
\nn\\
C&= \left\{
	\begin{array}{ll}
		r^2 -x + \sqrt{ (r^2 - x )^2-1} 
					&~~~~~{\rm when} ~~~~0<r<r_1 
							\quad {\rm or}~~ r> r_3  \\
		r^2 -x -i \sqrt{ 1-(r^2 - x )^2}
					&~~~~~{\rm when}~~~ r_1<r <r_3 
 	\end{array} 
\right.
\label{BC}		
\end{align} 
where  $r_1= \sqrt{x-1}$, $r_2 =\sqrt{x}$, 
and  $r_3=\sqrt{x+1}$.

Noting that the one-loop correction 
$S_F^{(1)}(p)$ 
is proportional to the measure factor 
$e^{\alpha p^0 } $, 
one can shift the propagator function 
\begin{equation}
\Delta^{-1} ( p) \to \Delta^{-1} ( p) 
-  \Delta \Gamma_2^{(1)}(p) ,
\end{equation}
which will shift the mass 
$
m_\kappa^4 \to  m_\kappa^4 
+ 4\kappa^4 \lambda  \,{ F(x)}/x$ 
or equivalently
\begin{equation}
m^2  \to m^2 
+ \lambda \kappa^2 \, \frac{ F(x)}x\,.
\end{equation} 
It turns out
that $F(x)$ is finite but has imaginary part, 
\begin{align}
\frac{\pi^2} {\sqrt 2} 
{\rm Re} \left(  F( x) \right)  
&= -\int_0^\infty dr 
 \frac{ r^2 \ln(B^2)}{4\pi \sqrt{ (r^2 + x)^2-1}}  
+\Big(\int_0^{r_1}+\int_{r_3}^\infty \Big) dr   
 	\frac{ r^2 \ln(C^2)}{4\pi \sqrt{ (r^2 - x)^2-1}} 
\nn\\
&~~~ +\int_{r_1}^{r_3} dr 
\frac{ r^2 \cos^{-1}( r^2-x) } 
{2 \pi \sqrt{1-( r^2-x)^2}}  
-\int_{r_2}^{r_3}  dr
 \frac{r^2} 
{2 \sqrt{1-( r^2-x)^2}}  
\\
\frac{\pi^2} {\sqrt 2} 
{\rm Im } \left( F( x) \right)  
& = -\int_0^\infty  dr 	
\frac{ r^2} {2\sqrt{ (r^2 + x)^2-1}}
+\int_{r_3}^\infty  dr 	
\frac{r^2} {2\sqrt{ (r^2 - x)^2-1}}\,.
\end{align}

The quadratic divergence in the particle and 
anti-particle contribution at large momentum
is compensated by the tachyon contribution.
Explicit evaluation is given as 
\begin{equation}
F(x)= 0.01803 + i/{\pi^2} + O(x-1)\,.
\end{equation}
The price for this finiteness is that  
$F(x)$ is not real. 
The imaginary contribution  
arises from the complex poles 
present in the propagator,
which have the role in the off-shell 
loop correction. 

\begin{figure}[htbp]
\begin{center}
\includegraphics[height=70mm]{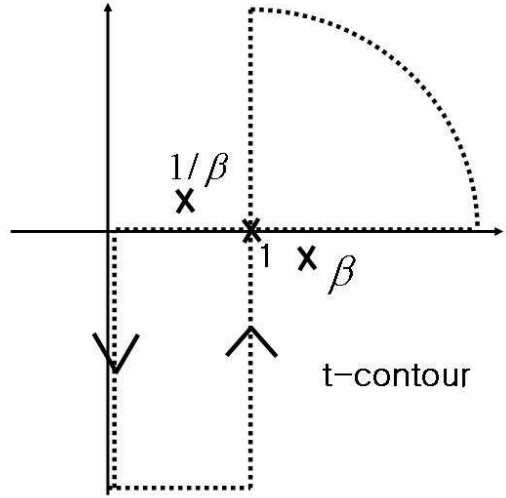}
\end{center}
\caption{Contour integration along $t=1$}
\label{fig:contour2}
\end{figure}

This imaginary contribution can be seen 
using a different contour integration of $t$
as in Fig.~\ref{fig:contour2}
where 
\begin{align}
& \int_0^\infty  
\frac{dt}{(t -\beta+ i\epsilon )
(t-1/\beta - i\epsilon )} 
\nn\\
&\quad 
=\int_{-\infty}^\infty   
\frac{id\tau}{(1+i\tau -\beta)
(1+i\tau-1/\beta  )} 
-\int_0^\infty 
\frac{id\tau}{(i\tau -\beta )
(i\tau-1/\beta  )} 
\nn\\ 
&\quad=\frac{1}{\beta - 1/\beta} 
\left( 
\ln \Big( \frac{1-\beta + i\tau}{ 1-1/\beta + i \tau} 
\Big)\Big|_{\tau=-\infty}^{\tau=\infty}  
- \ln \Big(\frac{i\tau-\beta }{i\tau-1/\beta } 
\Big)\Big|_{\tau=0}^{\tau=\infty} 
\right)  
\\
&\quad=\frac{1}{\beta - 1/\beta} \left( -{2\pi i}
- {\ln \beta^2 }  \right)
\nn
\end{align}
which is the same as the one 
in (\ref{IP}). 
The contour in Fig.~\ref{fig:contour2},
on the other hand, 
can be regarded as the Wick rotation 
$ p^0 \to i p^0$ at $\kappa \to \infty$,
since $t-1 \cong p^0/\kappa$. 
This shows that 
one cannot do the Wick rotation 
without including the poles at the complex $p^0$-plane 
due to the periodicity of $\Delta(p)$ in 
(\ref{periodicity}) 
in this KMST theory.

In addition, as $\kappa \to \infty$
the one loop correction becomes infinite
since $\Delta \Gamma_2^{(1)}(p)$ is quadratic 
in $\kappa$. 
This forces one 
to renormalize away 
the imaginary mass correction
as well as the quadratic $\kappa$ term
to have the proper theory 
at $\kappa \to \infty$ limit;
$m^2 + \Delta m^2$ with 
\begin{equation}
\Big(\Delta m^2\Big) ^{(1)}
=- \lambda \kappa^2 \, \frac{ F(x)}x\,.
\end{equation}

\section{one loop correction of the vertex} 
\label{sec:5} 
The one-loop correction of the 
four point function is given as
\begin{align}
\Gamma_4^{(1)} (p_1,p_2; p_3,p_4) 
=& i \lambda^2 \,\, e^{ \alpha(p_3^0+p_4^0)}\,
\int \frac{d^4 q}{(2\pi)^4} 
\left\{
{\Delta \Big(q+ \frac{p_1-p_3}2 \Big) \,
\Delta \Big(q- \frac{p_1-p_3}2 \Big) }
\right.
\nn\\
& ~~~~~~
\left. 
+ \frac 12 \Delta \Big( \frac{p_1 + p_2}2 + q \Big) 
\Big( \frac{p_1 + p_2}2 - q \Big) 
+ p_3 \longleftrightarrow p_4 
\right\}
\end{align}

\begin{figure}[htbp]
\begin{center}
\includegraphics[height=70mm]{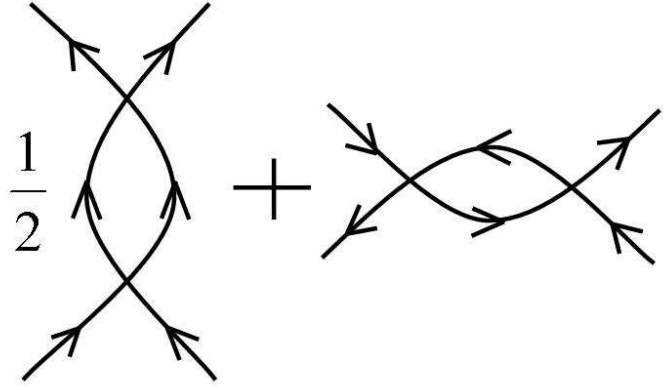}
\end{center}
\caption{One loop vertex correction}
\label{fig:oneloopv}
\end{figure}

At $p_i=0$ one has the one-loop correction,
$\Gamma_4^{(1)} $
as a function of $ x=m_\kappa^2 /(2\kappa^2) $
\begin{align}
\Gamma_4^{(1)} (0;x) 
&=  \lambda^2 
\int \frac{d^4 q}{(2\pi)^4}  \Delta^2 (q )
= \frac{3 \sqrt2}{\pi^2} \, \lambda^2 \int_0^\infty r^2 dr {\cal I}(r;x) 
\nn\\
{\cal I}(r;x) 
& = i\int_0^\infty \frac{  dt }{2\pi t} 
\Big( \frac{4t^2}
{(t-B)(t-1/B)
(t-C)(t-1/C)} \Big)^2
\\
&=i
\int_0^\infty \frac{ dt}{2\pi} \,\, t \,   
\Big( \frac{1}{(t-B)(t-1/B)}
- \frac{1}{(t-C)(t-1/C)} \Big)^2  
\nn\\
&= {\cal I}_C(r;x) +  {\cal I}_W(r;x) 
\nn
\end{align}
where $r =| {\bf q}|/(\sqrt2\kappa^2) $
and $B$ and $C$ are defined in (\ref{BC}).
$ {\cal I}_C(r;x)$ is 
the imaginary axis contribution 
\begin{align}
{\cal I}_C(r;x) 
&= 
i\int_0^\infty \frac{  d\tau }{2\pi \tau} 
\Big( \frac{4\tau^2}
{(i\tau-B)(i\tau-1/B)
(i\tau-C)(i\tau-1/C)} \Big)^2
\end{align}
and $ {\cal I}_W(r;x) $ is the pole contribution,
\begin{align}
 {\cal I}_W(r;x) &=
-\theta(r) 
\left( \frac{B^2(1+B^2)}{(B^2-1)^3}
+ \frac1{2x} \frac{B^2}  {B^2-1}
\right) 
\nn\\
&
-\theta(r-1)    
\left( \frac{C^2(1+C^2)}{(C^2-1)^3}
- \frac1{2x} \frac{C^2}  {C^2-1} \right)
\end{align}    

Let us consider the value at the imaginary axis.
One may put $ {\cal I}_C (r) $ in a more convenient 
form,  
\begin{align} 
{\cal I}_C (r;x)&=
i\int_0^\infty \frac{  d\tau }{2\pi \tau} 
 \frac{1}
{( r^2+ x-1+ a_0)^2 ( r^2+ x-1 -a_0^*)^2  }   
\end{align}
where 
$a_0 = 1 - i(\tau-1/\tau)/2$
and $a_0^*$ is the complex conjugate of 
$a_0$. 
Introducing $\tau = e^\theta$, 
one has $a_0 = \cosh \theta \, e^{-i \xi}$
and $\tan \xi \equiv\sinh \theta $.
Integration of $ {\cal I}_C (r;x) $ 
over $r$ is given as 
\begin{align}
{\cal C}(x) &= 
\int_0^\infty r^2 dr {\cal I}_C (r;x) 
\nn\\
&= i\int_{-\infty}^\infty \frac{  d\theta }{2\pi} 
\int_0^\infty dr 
 \frac{ r^2}
{( (r^2+x-1)^2+ (a_0-a_0^*)( r^2+x-1) -1)^2} \,.
\end{align}
After this, one may interchange the 
integration of $r$ and $\tau$. 
Integration over $r$ 
at the massless limit ($x=1$)
is given as  
\begin{align}
\int_0^\infty dr 
 \frac{ r^2}
{( r^2+ a_0)^2 ( r^2-a_0^*)^2  } 
&= \frac1{(\cosh\theta)^{5/2}} \int_0^\infty ds
 \frac{ s^2}
{( s^4 - 2 i( \sin \xi )s^2 +1)^2  } 
\nn\\
\int_0^\infty ds
 \frac{ s^2}
{( s^4 - 2 i( \sin \xi )s^2 +1)^2  } 
&=\left\{ 
\begin{array}{ll}
 \frac{\pi e^{3i\xi/2}}{4 (e^{i\xi} +i)^3} & 
 ~~~~{\rm  when} ~~ \theta>0  \\
 \frac{\pi e^{3i\xi/2}}{4 (e^{i\xi} -i)^3} & 
 ~~~~{\rm  when} ~~  \theta <0 
 \end{array}
\right.  \,.
\end{align}
And the integration over $\tau$ gives 
\begin{align}
 {\cal C}(x=1)
&= 
i\int_{-\infty}^\infty \frac{  d\theta}{2\pi } 
 \frac1{(\cosh\theta)^{5/2}}
\left(
\theta(\theta)  \frac{\pi e^{3i\xi/2}}{4 (e^{i\xi} +i)^3}
+\theta(-\theta)  \frac{\pi e^{3i\xi/2}}{4 (e^{i\xi} -i)^3}
\right)
\nn\\
&=\frac{i}8 \int_{-\infty}^\infty 
 \frac{d\theta}{\cosh\theta}
\left(
\theta(\theta)  
\frac{(1 + i \sinh \theta )^{3/2}}  
{(1+ i (\sinh \theta + \cosh \theta))^3 }
 + \theta(-\theta)\,  
\frac{(1 + i \sinh \theta )^{3/2}}  
{(1+ i (\sinh \theta - \cosh \theta))^3 }
\right)
\nn\\
&= \frac{i}8 \left( \int_0^\infty \frac{d\theta}{\cosh\theta}
  \frac{(1 + i \sinh \theta )^{3/2}}  
{(1+ i e^\theta)^3 }
+ c.c.\right)
\nn\\
&= -0.033257 i
\end{align}
which is finite but imaginary, 
absent in the ordinary field theory at $\kappa \to \infty$. 
$x$ dependence of the integration is given in  
powers of $(x-1)$ or $(m/2\kappa)^2$.

\begin{figure}[htbp]
\begin{center}
\includegraphics[width=70mm]{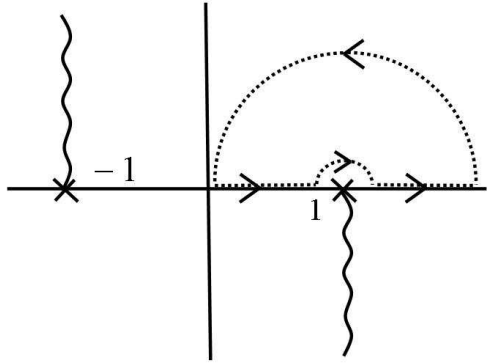}
\end{center}
\caption{contour integration along the upper 
half unit circle.}
\label{fig:bcontour}
\end{figure}

${\cal I}_W$ is conveniently rewritten in terms of $a=r^2+x$ and $b=r^2-x=a-2x$ 
using (\ref{BC}).:
\begin{align}
 {\cal I}_W(r;x) &= 
-\theta(a-x) 
\left( \frac{a}{4(a^2-1)^{3/2}} 
+ \frac1 {4x(a^2-1)^{1/2}}
\right)
\nn\\
&~~~~~~~~
-\theta(b) 
\left( \frac{b}{4(b^2-1)^{3/2}} 
-  \frac1 {4x(b^2-1)^{1/2}}
\right) 
\end{align}   
Integrating ${\cal I}_W(r;x) $ over $r$,
one has  
\begin{align}
\label{IWintegration}
\int_0^\infty 
 dr r^2 {\cal I}_W(r;x) 
& = - \frac18\int_x^\infty 
da \sqrt{a-x}   
\left( \frac{a}{(a^2-1)^{3/2}} 
+ \frac1 {x(a^2-1)^{1/2}}
\right) 
\nn\\
&- \frac18\int_0^\infty 
db \sqrt{b+x} 
\left( \frac{b}{(b^2-1)^{3/2}} 
-  \frac1 {x(b^2-1)^{1/2}}
\right) 
\end{align} 
In this way, one meets the role of 
branch cut at $b=\pm 1$
given in Fig.~\ref{fig:bcontour}. 
For example, 
one may evaluate the integration of $b$
avoiding the branch cut 
along the upper half unit circle in 
Fig.~\ref{fig:bcontour};
\begin{align}
\int_0^2 \frac{db}{(b-1)^{n/2}}
&= \int_0^{1-\epsilon} \frac{db}
{\left( e^{-i\pi} (b-1) \right)^{n/2}} 
+ \int_{1+\epsilon}^2 \frac{db}{(b-1)^{n/2}}
-\int_0^{\pi} \frac{i\epsilon e^{i\phi} d\phi}
{\left(\epsilon e^{i\phi}\right)^{n/2}} 
\nn\\
\label{bcontour}
&= i\int_{-\pi}^0 \, d\phi \,e^{i(1-n/2)\phi}
=\frac2{n-2} \left(
-e^{i\pi n/2} -1 
\right)
\end{align}
which has no $\epsilon$ dependence in the final result
for integer $n$.

One can conveniently divide 
the integration~(\ref{IWintegration}) as follows:  
\begin{align}
\int_0^\infty 
 dr r^2 {\cal I}_W(r;x) 
=  {\cal W}_1 (x)
+ {\cal W}_2 (x)
+  {\cal W}_3 (x) \,.
\end{align}  
${\cal W}_1 (x) $ is the ultraviolet contribution
\begin{align}
{\cal W}_1 (x) 
&=  - \frac18\int_2^\infty d c    
\left( \frac{c( \sqrt{c-x}+ \sqrt{c+x})}{(c^2-1)^{3/2}} 
+ \frac{\sqrt{c-x}-\sqrt{c+x}} {x(c^2-1)^{1/2}}
\right) 
\nn\\
&= -0.199907 + O(x-1)
\end{align} 
which turns out to be finite. 
${\cal W}_2 $ and  ${\cal W}_3$ are 
infrared parts. 
${\cal W}_2$ involves the branch cut 
at $b= \pm 1$ in Fig.~\ref{fig:bcontour}:
\begin{align}
{\cal W}_2 (x)
&=- \frac18\int_0^2 db \, 
\frac{\sqrt{b+x}} {(b^2-1)^{3/2}} 
\left( b- \frac{b^2-1} {x}
\right) \,.
\end{align}
Putting $b= 1 + e^{i \phi}$ one has
\begin{align}
{\cal W}_2 (x)
&= -\frac{i}8 \int_{-\pi}^0  
d \phi \, \frac{e^{-i\phi/2} \sqrt{1+ x +e^{i \phi}}  } 
 {(2 + e^{i \phi})^{3/2} } 
  \left(  
 1 + e^{i \phi} -  \frac{e^{i \phi}(2 + e^{i \phi})} {x}
 \right) 
\nn\\
&=0.320599 +  0.0470968 i + O(x-1)\,.
\end{align}
${\cal W}_3 (x) $, on the other hand, 
is infrared sensitive 
\begin{align}
{\cal W}_3 (x)
&= - \frac18\int_x^2 
da \sqrt{a-x}   
\left( \frac{a}{(a^2-1)^{3/2}} 
+ \frac1 {x(a^2-1)^{1/2}}
\right) 
\nn\\
&=\frac 1{16 \sqrt{2}} \ln(x-1)
-0.086266 + O(x-1)\,.
\end{align}
Note that the logarithmic term 
is divergent 
when $x \to 1 $ limit,
which can be considered either 
as the massless limit
or as $\kappa \to \infty$ limit. 
Combining all the terms, one has 
\begin{align}
\Gamma_4^{(1)} (0;x) 
&=
 \frac{3 \sqrt2}{\pi^2} \lambda^2 
 \left( \frac 1{16 \sqrt{2}} \ln(x-1)  +
 0.034426 + 0.013840 i 
 +  O(x-1)
\right)\,.
\end{align}

\section{Summary and discussion} 
\label{sec:6} 

We considered an interacting 
complex scalar field theory in
the $\kappa$-Minkowski spacetime. 
The theory is given in momentum space representation
based on the symmetric ordering of the 
$\kappa$-deformation of Poincar\'e algebra. 

Explicit calculation shows that the 
one loop correction is finite,
which has been the old dream of 
of non-commutative theory  
since Snyder's work \cite{snyder}.
Even though the theory is finite, 
the theory does show the divergent behavior
as $\kappa \to \infty$ limit 
since the propagator correction 
is order of $\kappa^2$
and the vertex correction  
is order of $\ln(m/\kappa)$.
In addition, 
the loop correction inevitably induces 
the imaginary correction 
due to the presence of the complex poles 
present in the propagator.
Thus, one needs to make a finite renormalization
to have the ordinary complex field theory 
at the $\kappa \to \infty$ limit.
In this way, one can see 
a renormalization group flow 
of the theory in terms of 
the new scale $\kappa$.
It is worth to mention that the logarithmic
dependence of $\kappa$ appears 
in the infrared sensitive way 
through the ratio of Planck scale $\kappa$ 
and the infrared scale $m$.  

Finally, one can confirm that 
$\kappa$-deformed Poincar\'e symmetry 
is respected in this complex scalar theory
even after the loop correction
since the exponential measure factor 
is maintained.   
This is because 
in this A-type interaction (\ref{A-type}) 
the exponential measure term in the vertex 
and the one in the internal propagator compensate 
each other as far as the  internal momentum 
is concerned.
If one considered a real scalar theory,
then the Bose symmetry
requires the measure factor 
to be a cosh function
rather than an exponential  
as in $B$-type interaction (\ref{B-type}) 
and this would spoil the KPS 
after the loop correction.

\begin{acknowledgments}
This work was supported in part 
by the Korea Research Foundation Grant funded by
Korea Government 
(MOEHRD, Basic Research Promotion Fund, KRF- C00153) and  
by the the Center for Quantum Spacetime (CQUeST)
of Sogang University with grant (R11-2005-021).
It is also acknowledged that 
the author has been benefited from the discussion 
with Prof. J.~Yee and 
the part of this article has been 
completed during his visit to 
Korea Institute for Advanced Study (KIAS).
\end{acknowledgments} 

\end{document}